\documentclass[RNAAS]{aastex631}

\usepackage{CJK}

\begin{document}
\begin{CJK*}{UTF8}{bsmi}

\title{$K_s$-band photometry of the Extreme T Subdwarf CWISE\,J221706.28$-$145437.6 \footnote{Based on observations made with the Gran Telescopio Canarias.}}

\author[0000-0001-5392-2701]{Jerry J.-Y.\ Zhang (章俊龑)}
\affiliation{Instituto de Astrof\'isica de Canarias (IAC), 
Calle V\'ia L\'actea s/n, E-38200 La Laguna, Tenerife, Spain}\affiliation{Departamento de Astrof\'isica, Universidad de La Laguna (ULL), E-38206 La Laguna, Tenerife, Spain}

\author[0000-0002-3612-8968]{Nicolas\ Lodieu}
\affiliation{Instituto de Astrof\'isica de Canarias (IAC), 
Calle V\'ia L\'actea s/n, E-38200 La Laguna, Tenerife, Spain}\affiliation{Departamento de Astrof\'isica, Universidad de La Laguna (ULL), E-38206 La Laguna, Tenerife, Spain}

\author[0000-0002-1208-4833]{Eduardo L. Mart\'in}
\affiliation{Instituto de Astrof\'isica de Canarias (IAC), Calle V\'ia L\'actea s/n, E-38200 La Laguna, Tenerife, Spain}\affiliation{Departamento de Astrof\'isica, Universidad de La Laguna (ULL), E-38206 La Laguna, Tenerife, Spain}

\correspondingauthor{Jerry Zhang}
\email{jzhang@iac.es}



\begin{abstract}
We present deep $K_s$-band imaging of the extreme T subdwarf CWISE\,J221706.28$-$145437.6. Using the new photometry, we construct its spectral energy distribution and find this object exhibits exceptionally strong collision-induced absorption in the $H$ and $K$ band. The comparison with the nearest benchmark extreme T subdwarf WISEA\,J181006.18$-$101000.5 suggests the object would be cooler and more metal-poor than the benchmark.

\end{abstract}



%
\section{Introduction}
\label{esdT_WISE2217:intro}

CWISE\,J221706.28$-$145437.6 (WISE2217) was discovered by \citet{meisner2021esdT} in the Backyard Worlds: Planet 9 citizen science project \citep{kuchner2017backyard}. It was identified as one of the few extreme T subdwarf (esdT) candidates \citep{meisner2021esdT,zhangjerry2023optical_sdT} because of its large motion, infrared (IR) and optical-IR colors similar to those of two known esdTs WISEA\,J181006.18$-$101000.5 (WISE1810) and WISEA\,J041451.67$-$585456.7 \citep{Schneider2020W0414_W1810}. Its subdwarf nature was supported by ground-based parallax measurements and multi-band colors \citep{zhangjerry2025optical}.

Although no spectroscopy of WISE2217 has been obtained due to its faintness, its spectral energy distribution (SED) can still be reconstructed using available photometry. However, a crucial piece of the SED is missing: no $K$-band photometry has been reported to date.  This is particularly significant because flux beyond 1.5 microns is a key diagnostic for identifying metal-poor T dwarfs. These subdwarfs exhibit strong collision-induced absorption (CIA) between hydrogen molecules, and between helium atoms and hydrogen molecules in their dense, high-gravity atmospheres, which leads to substantial flux suppression in the $H$ and $K$ bands. Hence, $K$-band photometry is essential to further constrain the nature and metallicity of WISE2217 through its SED.


\section{Observations and Data Reduction}
\label{esdT_WISE2217:Obs}

We collected new $K_{s}$-band photometry of WISE2217 with the Espectrografo Multiobjeto Infra-Rojo \citep[EMIR;][]{garzon2022EMIR} mounted on the 10.4-m Gran Telescopio Canarias (GTC).
The data were taken on the second half of the night of 25 July 2024 (MJD 60517.61) in visitor mode under clear skies but with Calima, 0\farcs5 seeing, and a bright moon. We used a 7-point dithering pattern with a 10\arcsec offset, and a 5s individual exposure at each dithering position. The pattern was repeated for 40 times, yielding a total on-source exposure of 1400s.

We reduced the images with the official pipeline PyEMIR \citep{Cardiel2019pyemir}. 
We have a $\sim$\,5-$\sigma$ detection of WISE2217 based on its previous position \citep{zhangjerry2023optical_sdT} and well-constrained proper motion \citep{zhangjerry2025optical}. We performed aperture photometry using the {\tt{Photutils}} package \citep{photutil} using apertures with a radius of 0\farcs8, sky annuli with an inner and outer radius of 3\arcsec and 5\arcsec, respectively. The reference stars are from the VISTA Hemisphere Survey \citep[][]{mcmahon2013vhs,mcmahon2020vhs_dr5}. 


\begin{figure}[htbp]
    \centering
    \includegraphics[width=\linewidth]{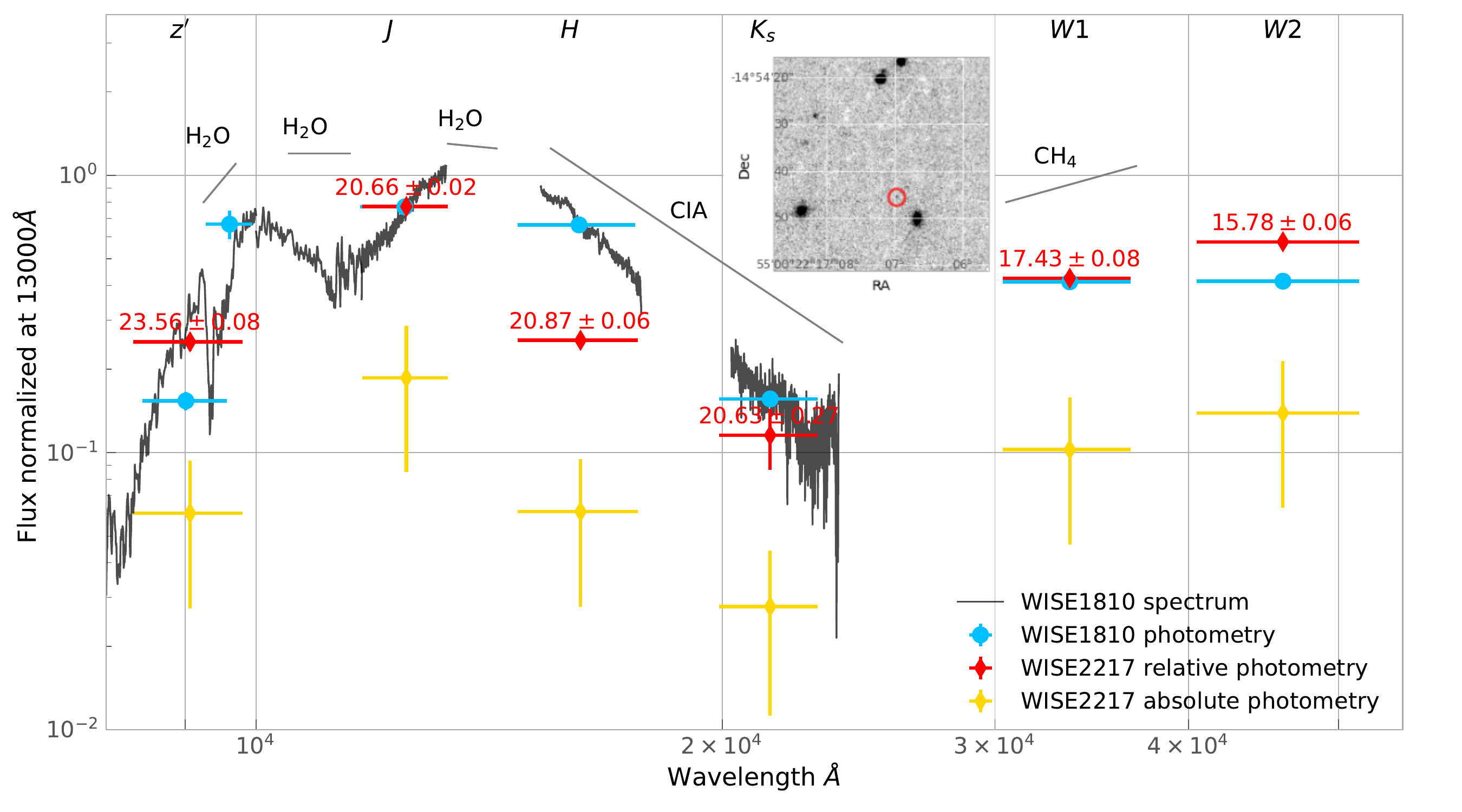}
    \caption{WISE2217's field (the small panel), photometry (red diamonds and magnitude values), with those of the benchmark esdT WISE1810 (blue circles) and its spectrum (black). The relative photometry of the two objects is aligned in the $J$ band. WISE2217's absolute photometry (yellow diamonds) is also scaled to WISE1810's $J$-band photometry. The uncertainty of the absolute photometry is mainly from the parallax uncertainty.}
    \label{photometry}
\end{figure}

\section{Result and Discussion}

We have a measurement of $K_s=20.63\pm0.27$\,mag (Vega). Fig~\ref{photometry} compiled all the photometry of WISE2217 in the optical \citep{zhangjerry2023optical_sdT} and IR \citep{meisner2023coldoldBD}.
We compared the SED of WISE2217 with that of the benchmark extreme T subdwarf WISE1810 (Figure~\ref{photometry}), which is the nearest T subdwarf \citep[8.9\,pc;][]{lodieu2022W1810} and has a well-constrained metallicity ($\mathrm{[M/H]}=-1.7\pm0.2$\,dex) derived from the NIR methane feature \citep{zhangjerry2025W1810}. In comparison, WISE2217 demonstrated a much stronger CIA in the $H$ and $K$ bands, indicating a more metal-poor atmosphere. We note that if we take into account the filter difference \citep{zhangjerry2023optical_sdT}, WISE2217 (Sloan $z'$) would have a similar $z-J$ color as WISE1810 (Pan-STARRS $z$). WISE2217 has a redder $W1-W2$ color. Considering that the methane absorption gets weaker in the $W1$ band when the metallicity lowers but gets stronger when temperature drops, as well as that the emission peak moves towards $W2$ band for T dwarfs when the temperature drops, a redder $W1-W2$ color infers a cooler temperature. This result supports the spectral type assignation for WISE1810 \citep[esdT0--esdT3;][]{Schneider2020W0414_W1810,burgasser2025esdT_class} and WISE2217 \citep[photometrically classified as esdT5.5$\pm$1.2,][]{meisner2021esdT}, if a good classification scheme for esdTs is a monotonic non-increasing function for the effective temperature.

WISE2217 has a parallax of $48\pm13$\,mas \citep{zhangjerry2025optical}, yielding 
that WISE2217's absolute magnitudes in all bands are fainter than those of WISE1810 by a confidence level of about 3$\sigma$. This lower luminosity could be explained by the synergy between the coldness and the low metallicity in its atmosphere.

In summary, WISE2217 could be the most metal-poor T dwarf up to date, as suggested by \citet{zhangjerry2025optical}. It is likely to have a metallicity [M/H] $\lesssim-2.0$\,dex and a temperature cooler than 1000\,K, by comparing with the benchmark esdT WISE1810. A follow-up with a high-quality NIR spectroscopy for WISE2217 is challenging but worthwhile, as we can use the most prominent NIR methane feature and the ATMO2020++ model \citep{leggett2021coldestSED,meisner2023coldoldBD} to precisely constrain its metallicity to 0.2\,dex or better.

%
%
\section*{acknowledgments}
JYZ and NL 
acknowledge support from the Agencia Estatal de Investigaci\'on del Ministerio de Ciencia, Innovación y Universidades under grant PID2022-137241NB-C41\@. 
JYZ and EGE were funded for this research by the European Union ERC AdG SUBSTELLAR grant agreement number 101054354\@.
Based on observations made with the Gran Telescopio Canarias (GTC), in the Spanish Observatorio del Roque de los Muchachos of the Instituto de Astrofísica de Canarias, on the island of La Palma, under program GTC37-23B (PI Zhang).
EMIR has been funded by GRANTECAN S.L.\ via a procurement contract; by the Spanish funding agency grants AYA2001-1656, AYA2002-10256-E, FIT-020100-2003-587, AYA2003-01186, AYA2006-15698-C02-01, AYA2009-06972, AYA2012-33211, AYA2015-63650-P and AYA2015-70498-C2-1-R; and by the Canarian funding agency grant ACIISI-PI 2008/226.

\vspace{5mm}
\facilities{GTC}

\software{
PyEMIR\footnote{\url{https://pyemir.readthedocs.io/}}, 
Photutils \citep{photutil}, Astropy \citep{astropy2013,astropy2018,astropy2022}}

\bibliography{bibliography}{}
\bibliographystyle{aasjournal}


\end{CJK*}
\end{document}